\documentclass[aps,showpacs,twocolumn]{revtex4}
\usepackage[dvips]{graphicx}
\usepackage{amsmath,array}
\usepackage{color}
\usepackage{natbib}
\usepackage{wrapfig}
\newcommand{\cdlq}{c_{\text{L}q}^{\dagger}}
\newcommand{\clq}{c_{\text{L}q}}

\newcommand{\dd}{d^{\dagger}}

\newcommand{\elq}{\varepsilon_{\text{L}q}}
\newcommand{\epo}{\varepsilon_0}
\newcommand{\G}{\Gamma}
\newcommand{\la}{\langle}
\newcommand{\ma}[1]{{\mathcal #1}}
\newcommand{\ra}{\rangle}
\newcommand{\tl}{{\rm L}}
\newcommand{\tr}{{\rm R}}
\newcommand{\vlq}{V_{\text{L}q}}

\begin{document}

\title{Electronic Heat Transport Across a Molecular Wire:\\
Power Spectrum of Heat Fluctuations}

\author{Fei Zhan, Sergey Denisov, and Peter H\"anggi}
\affiliation{Institut f\"ur Physik, Universit\"at Augsburg,
Universit\"atsstr.~1, D-86159 Augsburg, Germany}

\begin{abstract}
With this study we analyze the fluctuations of an electronic  heat current across a molecular wire. The wire  is composed of  a single energy level which connects two  leads which  are held at different temperatures. By use of the Green function method we derive an explicit expression for the  power spectral density of the emerging heat noise.  This result  assumes a form that is quite  distinct from the power spectral density  of the accompanying electric current noise. The complex expression simplifies considerably in the limit of zero frequency, yielding the heat noise intensity.  The heat noise spectral density still depends  on the frequency in the zero-temperature limit, assuming different asymptotic behaviors  in the  low- and high-frequency regions. These findings evidence that  heat transport across molecular junctions can exhibit a  rich structure beyond the common behavior which emerges in the  linear response limit.
\end{abstract}
\date{\today}

\maketitle

\section{Introduction}
The experimental realization of molecular junctions~\cite{reed97science} some fifteen years ago has sparked a wave of research activities both in theoretical \cite{blanter00,kohler05,dubi11rmp} and the experimental \cite{joachim00nature,tao,cui01science,reichert02prl}
communities. Presently, single molecule electronics is considered as a possible potential substitution of the silicon-based elements in
the information processing  technology~\cite{joachim00nature,tao}, and this perspective makes  studies of molecular transport properties very appealing. Besides the standard current-voltage characteristics~\cite{reed97science,cui01science,reichert02prl}, it is also  possible, for example by use of the full counting statistics~\cite{levitov04prb,morten08prb,bagrets03prb}, to obtain the information about the fluctuations of the electric current flowing through a molecular wire \cite{clement07prb,beenakker03,liyuanp90apl,buttiker92prb,camalet04prb}.

The issue of heat transport across such molecular junctions attracted much less attention in the prior literature. This may be so because a measurement of heat flow   experimentally is far from being a straightforward task. The problem of heat current constitutes, however,  an important physical issue. This is so because the structural {\it stability} of any molecular structure depends sensitively on the heat flow which accompanies the inter-electrode charge transport.

With the systems of interest being of nanoscale the corresponding  heat noise  can also be quite large. This may be so even in situations where the average heat current is identically {\it vanishing} all together, as it is the case in thermal equilibrium. Moreover, the properties of  noise correlation features, or likewise, its frequency-dependent spectral properties and, as well, its zero-frequency power spectrum, are in no way directly related to the mean value of the heat flow itself.  It is thus of outmost interest to gain some insight into the size of heat noise in such molecular wire setups. In particular, it would be very useful to have analytical estimates for its power spectral density (PSD)  available, even at the expense that these may predominantly apply to idealized setups only.

Heat transport across conducting molecular wires, explicitly induced by a difference of the lead temperatures, $T_{\rm L}$ and $T_{\rm R}$, is a  phenomenon which involves electrons, phonons and their mutual interaction ~\cite{zhanfei09pre,sivan86prb,koch04prb,chen05prl,galperin06prb,galperin07prb,segal06prb,paulsson03prb,{galperin07jpcm},
 wangjiansheng08epjb,dubi11rmp}. Therefore, the definition of heat, carried through the wire, should be addressed with care, with the
need to  distinguish between heat transfer mediated by electrons and a one mediated by phonons.

The latter issue constitutes the  realm of {\it phononics} \cite{arXivphononics}, a promising novel research area which may lead to new elements, such as molecular thermal diodes, thermal transistors,  thermal logic gates, to name but a few ~\cite{arXivphononics,changcw06science,wanglei07prl,libaowen04prl,libaowen06apl}. Here also, the size of fluctuations in heat current does matter; this is so because those may well turn out to be deleterious to intended information processing tasks.

Energy transport mediated by electrons is a process which is related to its electric current: electrons are moving from lead to lead, carrying not only  charge but also energy. However, the amount of energy carried by a single electron, unlike to its charge,  is not constant  ~\cite{rey07prb,moskalets04prb}. In contrast to the studies that examine the average heat flow  much less attention, however, has been paid to the issue and impact of  \textit{fluctuations} of heat flow across various  nanoscale devices. In some prior work ~\cite{krive01prb} the heat transport through a ballistic quantum wire has been considered in the  Luttinger-liquid limit, by neglecting the discreteness of the wire's energy spectrum. In more recent publications, Refs.~\cite{sergi11prb, averin10prl}, the PSD of the heat current fluctuations has been derived within the scattering theory, under the assumption that the electrons  are transmitted (reflected) at the same rates, independently of their actual energies. The results of the last two papers are distinctive because they have shown that  the  heat noise exhibits a well-pronounced frequency dependence even in the zero-temperature limit.


With this work we consider the  {\it electronic} heat current that  proceeds across  a molecular wire composed of a  single energy level with the two leads held at a constant temperature difference. In doing so we shall neglect electron-phonon interactions and electron-electron interactions. Such a simplification can be justified in situations that involve a short wire. Then, the Coulomb interaction via a double occupancy shifts the energy far above the Fermi level so that its role for thermal transport can be neglected. Likewise, the electron dwell time is very short as compared to the electron-phonon relaxation time scale.

In contrast to  prior works \cite{sergi11prb, averin10prl}, however,  we here take into account the dependence of the transmission coefficient on its electron energies, and, within the nonequilibrium Green function approach \cite{wangjiansheng08epjb,kohler05}, derive an explicit  expression for the  PSD of the heat current fluctuations, $\tilde{S}^{\rm h}(\omega)$. With this result at hand we explore different regimes of electron transport and demonstrate that the heat noise in fact is quite distinct from its electric counterpart.

In Sec.~III, we evaluate by use of the Green function method the expression  for both, electronic heat current and its noise. We study analytically the distinct properties of the electronic heat noise and compare this result with the electronic current noise counterpart. In doing so we demonstrate that for the chosen setup of a  molecular junction with a single molecular energy level  different power laws for the PSD occur in different frequency regions. We discuss the obtained findings and end with conclusions in Sec. IV.

\begin{figure}[b]
 \includegraphics[width=5cm]{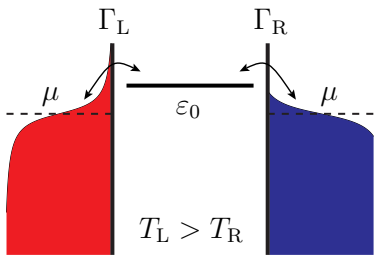}
\caption{\label{fig:model}(color online)~Setup of a molecular junction: two metal leads, filled with electron gas, are connected by a single orbital $\epo$. The coupling strengths are determined by the constants $\G_{\text{L/R}}$. The left lead is prepared at a higher temperature as compared to   the opposite  right lead, i.e.  $T_{\text{L}} > T_{\text{R}}$. The chemical potential, $\mu$, is the same for both leads so that no electric current due to a voltage bias is present.}
\end{figure}

\section{Model setup}
The molecular junction setup is depicted with Fig.~\ref{fig:model}: It  is  described by a Hamiltonian
\begin{equation}
 H=H_{\text{wire}}+H_{\text{leads}}+H_{\text{contacts}}\;, \label{totalham}
\end{equation}
containing three different contributions, namely the wire Hamiltonian,  the role of  leads and the wire-lead coupling, respectively. We consider here the regime of {\it coherent} quantum transport whereby  neglecting dissipation inside the wire. The wire is composed of a single orbital; i.e.,
\begin{equation}
H_{\rm wire}=\epo\dd d\;,\label{hammol}
\end{equation}
at an  energy $\epo$, with the fermionic creation and annihilation operators,  $\dd$ and $d$. The energy level $\epo$ can be tuned by applying a gate voltage. This idealized setup makes possible  explicit analytical calculations. It mimics a  double barrier resonant tunneling structure $\text{GaAs/Al}_x\text{Ga}_{1-x}$-structure of the type considered for electronic shot noise calculations in Ref. \cite{bo96jpcm}, herein truncated to a single Landau level. The leads are conventionally modeled by reservoirs, composed of ideal electron gases, i.e.,
\begin{equation}
H_{\text{leads}}=\sum_{\ell q}\varepsilon_{\ell q}c_{\ell q}^{\dagger}c_{\ell q}\;,\label{hamld}
\end{equation}
where the operator $c_{\ell q}^{\dagger}(c_{\ell q})$ creates (annihilates) an electron with momentum $q$ in the $\ell=$L (left) or $\ell=$R (right) lead. We assume that the electron distributions in the leads are described by the grand canonical ensembles at the temperatures $T_{\rm L/R}$ and with  chemical potentials $\mu_{\rm L/R}$. With such ideal electron reservoirs we then obtain $\la c_{\ell q}^{\dagger}c_{\ell' q'}\ra=\delta_{\ell\ell'}\delta_{qq'}f_{\ell}(\varepsilon_{\ell q})$, where
\begin{equation}
f_{\ell}(\varepsilon_{\ell q})=\left[e^{(\varepsilon_{\ell q}-\mu_{\ell})/k_{\text{B}}T_{\ell}}+1\right]^{-1}
\end{equation}
denotes the Fermi function.

Next we impose a finite  temperature difference $\Delta T=T_{\text{L}}-T_{\text{R}}$ and use identical  chemical potentials, $\mu_{\rm L}=\mu_{\rm R} = \mu$ for the leads. When an electron tunnels out from a lead, the energy $E$  leaks into the wire. This energy presents a heat contribution, $\delta Q$, which in terms of the chemical potential, $\mu$, reads $\delta Q = E-\mu$. In the following  we assume that  all the electron energies are counted from the chemical potential  value $\mu$, i.~e., we set $\mu=0$~\cite{zhanfei09pre,dubi11rmp}.

The tunnelling  Hamiltonian,
\begin{equation}
 H_{\text{contacts}}=\sum_{\ell q}V_{\ell q}c_{\ell q}^{\dagger}d + h.c.\;,\label{hamcon}
\end{equation}
mediates the coupling between the wire and the leads. The notation $h.c.$ denotes Hermitian conjugate.  The quantity
$V_{\ell q}$ is the tunnelling matrix element, and the tunneling coupling is characterized  in general by a  spectral density,
$\Gamma_{\ell}(E)=2\pi\sum_{q}|V_{\ell q}|^2\delta(E-\varepsilon_{\ell q})$ \cite{kohler05}. In what follows we also use the wide-band limit of the lead conduction bands, with  $\Gamma_{\ell}(E)=\Gamma_{\ell}$.

\section{Fluctuations of heat flow}
We perform the derivation of the central quantities, i.e. the heat current and its PSD of corresponding fluctuations,  within the Heisenberg description. Setting for the energy operator,
\begin{equation}
E_{\tl}=\sum_{q}\elq\cdlq\clq\;,
\end{equation}
its time derivative yields the operator for the heat current, reading:
\begin{equation}
J^{\rm h}_{\tl}(t)=-\sum_q\frac{2\elq}{\hbar}\text{Im}[\vlq\cdlq(t) d(t)]\label{jlop}\; .
\end{equation}
This current is  positive when heat transport proceeds from the left (i.e. hot) lead to the right (cold) lead, see in Fig. 1. Upon formally solving the Heisenberg equation of motion for the lead operators, we  obtain
\begin{align}
c_{\ell q}(t)=&c_{\ell q}(t_0)e^{-i\varepsilon_{\ell q}(t-t_0)/\hbar}\notag\\
&-\frac{iV_{\ell q}}{\hbar}\int_{t_0}^{t}dt'e^{-i\varepsilon_{\ell q}(t-t')/\hbar}d(t')\label{clq}\;,
\end{align}
where the first part on the right hand side describes the dynamics of the free electrons in the leads, while the second part accounts for the influence of the wire.

By inserting Eq. \eqref{clq} into the Heisenberg equation for the electron annihilation operator within the wire, we find
\begin{equation}
\dot{d}=\frac{i}{\hbar}\epo d-\frac{\G_{\rm L}+\G_{\rm R}}{2\hbar}d+\xi_{\text{L}}(t)+\xi_{\text{R}}(t)\;,\label{cem}
\end{equation}
where
\begin{equation}
 \xi_{\ell}(t)=-\frac{i}{\hbar}\sum_qV^*_{\ell q}\exp\left[-\frac{i}{\hbar}\varepsilon_{\ell q}(t-t_0)\right]c_{\ell q}(t_0)\;.
\end{equation}
To obtain the solution of Eq.~\eqref{cem}, we follow the Green function approach in Ref.~ \cite{kohler05}, and first solve the following differential equation
\begin{equation}
 (\frac{d}{dt}+\frac{i\varepsilon_0}{\hbar}+\frac{\G_{\rm L}+\G_{\rm R}}{2\hbar})G(t-t')=\delta(t-t')\;, \label{green}
\end{equation}
and then apply the convolution $d(t)=\int G(t-t')(\xi_{\tl}(t')+\xi_{\tr}(t'))dt'$.
The solution of Eq. \eqref{green} is given by:
\begin{equation}
 G(t)=\theta(t)e^{-i\epo t/\hbar-(\G_{\rm L}+\G_{\rm R})t/2\hbar}\;.
\end{equation}
Then the molecular operator in Eq.~(\ref{cem}) assumes the form
\begin{equation}\label{ct}
d(t)=\sum_{\ell q} V^*_{\ell q}\frac{\exp[-i\varepsilon_{\ell q}(t-t_0)/\hbar]}{\varepsilon_{\ell q}-\epo+i(\G_{\rm L}+\G_{\rm R})/2}c_{\ell q}(t_0)\;.
\end{equation}
Upon substituting this result into Eq. \eqref{clq}, we obtain for the operators in the leads
\begin{align}
&c_{\ell q}(t)=c_{\ell q}(t_0)e^{-i\varepsilon_{\ell q}(t-t_0)/\hbar}\notag\\
&+\sum_{\ell' q'}\frac{V_{\ell q}V^*_{\ell' q'}e^{-i\varepsilon_{\ell' q'}(t-t_0)/\hbar}}{\varepsilon_{\ell' q'}-\epo+i(\G_{\rm L}+\G_{\rm R})/2}c_{\ell' q'}(t_0)\notag\\
&\times B[\varepsilon_{\ell'q'}-\varepsilon_{\ell q}]\;,\label{clqt}
\end{align}
where,
\begin{equation}
B(E)={\mathcal P}\left(\frac{1}{E}\right)-i\pi\delta(E)\;,
\end{equation}
and  ${\mathcal P}$ denotes the integral principal value \cite{soh}.

Next we  insert Eq.~\eqref{ct} and Eq.~\eqref{clqt} into the heat current operator, Eq.~\eqref{jlop}, and by consequently taking  the ensemble average, we obtain a Landauer-like formula for the heat current~\cite{segal03jcp,dubi11rmp,galperin07jpcm}; reading,
\begin{equation}\label{current}
 \la J^{\rm h}(t)\ra := J^{\rm h} = \frac{1}{2\pi\hbar}\int dEE{\mathcal T}(E)[f_{\tl}(E)-f_{\tr}(E)],
\end{equation}
where the transmission coefficient, $\ma{T}(E)=\G_{\tl}\G_{\tr}/[(E-\epo)^2+\G^2]$, is  energy-dependent. Below we consider the case of symmetric coupling between the wire and the leads,  $\G_{\rm L}=\G_{\rm R}=\G$.

\begin{figure}[t]
{\center\includegraphics[width=8cm]{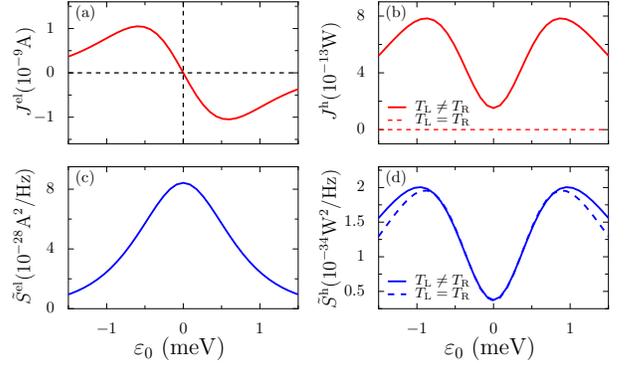}}
\caption{\label{fig:curnoise} (color online) Currents (top row) and zero-frequency components of PSDs  of accompanying noises (bottom row) for charge (left column) and  heat (right column) transport through the single-orbital wire as  functions of orbital energy for $\Gamma=0.1$~meV. The remaing parameters are $T_{\rm L}=5.2$~K, $T_{\rm R}=3.2$~K (solid lines). For  equal chemical potentials $\mu_{\rm L}=\mu_{\rm R}=0$ the heat flow in panel (b) vanishes for equal temperatures   $T_{\rm L}=T_{\rm R}$; its noise intensity in equilibrium  at the temperature $T_{\rm L} = T_{\rm R} = 4.2$~K is depicted versus orbital energy $\varepsilon_0$  by the dashed line in panel (d).}
\end{figure}

The expression for the electric Seebeck current~\cite{dubi11rmp} reads very similar to Eq.~(\ref{current}), except for its absence of the energy multiplier $E$ in the integral in the rhs of Eq.~(\ref{current}). This seemingly small difference changes, however,  the  physics of the transport through the wire, because the multiplier inverts the symmetry of the integral. Namely, the Seebeck current is an antisymmetric function of orbital energy and vanishes when the orbital energy level is aligned to the  chemical potentials of the leads,  Fig.~\ref{fig:curnoise}(a), while the heat current is a symmetric function and acquires a nonzero value at $\varepsilon_0 = 0$, Fig.~\ref{fig:curnoise}(b).

The heat noise is described by the symmetrized autocorrelation function, i.e.,
\begin{equation}
S^{\rm h}(\tau)=1/2\la[\Delta J^{\rm h}_{\ell}(\tau),\Delta J^{\rm h}_{\ell}(0)]_+\ra\;,
\end{equation}
of the heat current fluctuation operator $\Delta J^{\rm h}_{\ell}(t)=J^{\rm h}_{\ell}(t)-\la J^{\rm h}_{\ell}(t)\ra$, where the anti-commutator  $[A,B]_+=AB+BA$ ensures the hermitian property.

In the asymptotic limit $t \rightarrow \infty$, the auto-correlation function   depends  on the time difference only.  Its Fourier transform is the {\it power spectrum} for heat noise, obeying
\begin{equation}\label{ss}
 \tilde{S}^{\rm h}(\omega) = \tilde{S}^{\rm h}(-\omega) = \int_{-\infty}^{\infty} d\tau e^{i\omega\tau}S^{\rm h}(\tau) \geq 0\;,
\end{equation}
being an even function in frequency and strictly semi-positive (Wiener-Khintchine theorem).
In the following we address  positive values of the frequency, $\omega > 0$, only.

\subsection{The spectrum of heat fluctuations: explicit results}
Upon combining  Eq.~\eqref{ss}  and Eq.~\eqref{jlop}, a cumbersome evaluation then yields the following, nontrivial  explicit expression for the PSD of electronic heat noise,
reading:
\begin{widetext}
\begin{align}
&\tilde{S}^{\rm h}(\Omega= \hbar\omega; T_{\tl},T_{\tr})\notag\\
=&\sum_{\pm}\int\frac{dE}{4\pi\hbar}\left\{\left[\left(E\pm\frac{\Omega}{2}\right)^2\ma{T}(E)\ma{T}(E\pm\Omega)+\frac{\G_{\rm L}^2\left[E(E-\epo)-(E\pm\Omega)(E\pm\Omega-\epo)\right]^2}{\left[(E-\epo)^2+\G^2\right][(E\pm\Omega-\epo)^2+\G^2]}\right]\right.f_{\rm L}(E)\overline{f}_{\rm L}(E\pm\Omega)\notag\\
&+\left(E\pm\frac{\Omega}{2}\right)^2\ma{T}(E)\ma{T}(E\pm\Omega)f_{\rm R}(E)\overline{f}_{\rm R}(E\pm\Omega)+[E^2\ma{R}(E)\ma{T}(E\pm\Omega)\mp \frac{1}{2}E\Omega\ma{T}(E)\ma{T}(E\pm\Omega)\notag\\
&+\left(E\pm\frac{\Omega}{2}\right)\left(\pm\frac{\Omega}{2}\right)\frac{\G{\rm L}^2\ma{T}(E\pm\Omega)}{(E-\epo)^2+\G^2}]f_{\rm L}(E)\overline{f}_{\rm R}(E\pm\Omega)+[\left(E\pm\Omega\right)^2\ma{R}(E\pm\Omega)\ma{T}(E)\notag\\
&+\left(E\pm\Omega\right)\left(\pm\frac{\Omega}{2}\right)\ma{T}(E)\ma{T}(E\pm\Omega)+\left(E\pm\frac{\Omega}{2}\right)\left(\mp\frac{\Omega}{2}\right)\frac{\G_{\rm L}^2\ma{T}(E\pm\Omega)}{(E-\epo)^2+\G^2}]f_{\rm R}(E)\overline{f}_{\rm L}(E\pm\Omega)\;,\label{longeqn}
\end{align}
\end{widetext}
wherein we abbreviated  $\Omega\equiv\hbar\omega$, $\overline{f}\equiv1-f$, and $\ma{R}(E)\equiv1-\ma{T}(E)$ is the reflection coefficient.

\begin{figure}[t]
{\center\includegraphics[width=8cm]{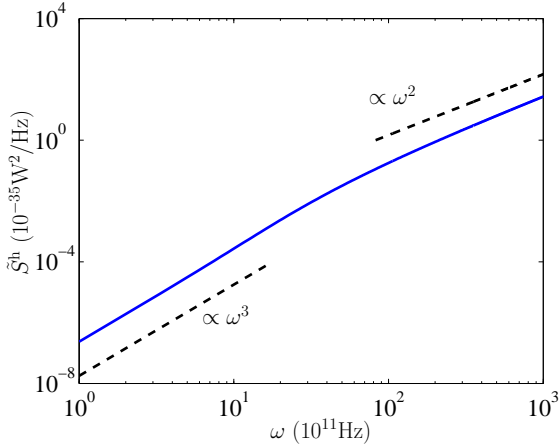}}
\caption{\label{fig:ze}(color online) At zero-temperature $T_{\rm L}=T_{\rm R}=0$ the dependence of heat noise  PSD versus frequency $\omega$ exhibits  different  power-law behaviors (dashed lines) for $\omega$ sampling intermediate values (proportional to $\omega^3$) as compared to the large frequency limit (proportional to $\omega^2$) when   $\omega \rightarrow \infty$.  The other parameters are  $\varepsilon_0=0$ and $\Gamma=0.1~{\rm meV}$.}
\end{figure}

The PSD of heat noise at zero frequency $\omega=0$ simplifies considerably, assuming an appealing form; reading
\begin{align}\label{hnoise}
&\tilde{S}^{\rm h}(\omega=0; T_{\tl},T_{\tr})\notag\\
=&\frac{1}{2\pi\hbar}\int dE E^2 [{\mathcal T}(E)[ f_{\tl}(E)[1-f_{\tl}(E)]\notag\\
&+f_{\tr}(E)[1-f_{\tr}(E)]]\notag\\
&+{\mathcal T}(E)[1-{\mathcal T}(E)][f_{\tl}(E)-f_{\tr}(E)]^2]\;.
\end{align}
This main result is in  agreement with a  conjectured prediction  in  Ref.~\cite{krive01prb}.
The distinct difference between Eq.~\eqref{hnoise} and the PSD of the fluctuations of the nonlinear, accompanying Seebeck electric current,
reading ~\cite{blanter00,kohler05},

\begin{align}
&\tilde{S}^{\rm el}(\omega=0; T_{\tl},T_{\tr})\notag\\
=&\frac{e^2}{2\pi\hbar}\int dE [{\mathcal T}(E)[ f_{\tl}(E)[1-f_{\tl}(E)]\notag\\
&+f_{\tr}(E)[1-f_{\tr}(E)]]\notag\\
&+{\mathcal T}(E)[1-{\mathcal T}(E)][f_{\tl}(E)-f_{\tr}(E)]^2]\;,
\end{align}
is a factor $E^2$ in the integral in Eq.~(\ref{hnoise}). Although this distinction seemingly appears minor, it leads to a tangible difference in a dependence of the noise PSDs on energy level of the wire orbital, as depicted with Fig.~\ref{fig:curnoise}. When the orbital level, $\varepsilon_0$,  is tuned to the chemical potential, the two expressions reveal  different properties: while the zero-frequency component  of the electric PSD at $\omega=0$ exhibits a maximum at $\varepsilon_0 = 0$, its heat counterpart possesses a local minimum at this value.

This difference originates  from the salient fact that the two  transport mechanisms for charge and the energy are not equivalent. The electric current  is quantized by the electron charge, $\overline{e}$; in contrast, the energy carried by the electron is not quantized and may assume principally an arbitrary value. The main contribution to the electron flow across the wire  stems from  the electrons occupying energy levels around the chemical potential $\mu$.

Because the  interaction between the leads and the molecule is weak, it only slightly perturbs the Fermi distributions, which possess strongly nonuniform profiles around $\mu=0$. Electrons of different energies contribute differently to the heat transport, but the Fermi distribution allows only for a finite number of electrons per energy level: i.e. just one in case of spinless electrons, or two in the case of electron with spin. Therefore, the electrons of given energy can move across the wire  only when the corresponding level of the destination lead can host them.

When $\varepsilon_0$ deviates from the chemical potential, increasingly less electrons  participate in the transport. The flow of electrons becomes diminished, and since both, the electric current and the electric noise are  insensitive to the electron energies, they both decrease with increasing  $|\varepsilon_0|$. This scenario differs for  heat transport: this is so because the deviation from the chemical potential  increases the possibility that successive electrons will carry different energies.  This in turn  will lead to an  increase of   heat noise. With  further deviation of the orbital energy from the chemical potential, the occupancy difference decreases monotonically and consequently the heat noise strength decreases again.

\subsection{Equilibrium heat fluctuations at $T_{\rm L} = T_{\rm R}$}

Next let us focus on the equilibrium properties of heat noise; i.e., the situation when the two temperatures are equal, $T_{\rm L}=T_{\rm R}$. In this case the average heat flow vanishes identically, but not its fluctuations. The zero-frequency spectra of  both noises, for heat and electric noise, i.e. the corresponding power spectra  increase with the increase of the coupling $\Gamma$, since it increases the transmission probability. The noise intensities are different from zero, however, even at equilibrium, see for heat noise Fig.~\ref{fig:curnoise}, i.e., the two panels (b,d).

The properties at vanishing temperature, $T_{\rm L}=T_{\rm R}=0$, are more subtle. It has been pointed in Ref.~\cite{averin10prl} that heat noise  seemingly  violates the fluctuation-dissipation theorem (FDT). In fact, a temperature difference, $\triangle T$, does not induce any thermal gradient. Therefore, there is no force which is conjugate  to the heat current, and the process is out of the validity region of the FDT. This `evasion' of the FDT is fully active in the zero-temperature limit, $T_{\rm L} = T_{\rm R} = 0$, where the heat noise PSD still depends on the frequency. This dependence is due to quantum fluctuations, the virtual transitions of electrons directly from lead- to-lead \cite{averin10prl,sergi11prb}.  The Fermi distribution equals the Heaviside step function in this case. Thus, the contributions to the integrand in Eq. \eqref{longeqn} comes from  the interval $[-\Omega,~0]$ only. After the integration of Eq. \eqref{longeqn}, one finds the central result for the frequency dependent PSD:
\begin{align}
&\tilde{S}^{\rm h}(\omega,T_{\tl}=T_{\tr}=0)\notag\\
=&\frac{\G}{4\pi\hbar}\left\{\left[(2\Omega)^2-2\G^2\right]\arctan\left(\frac{\Omega}{\G}\right)\right.\notag\\
&+2\left.\Omega\G\left[1+\log\left(\frac{\G^4}{\left(\Omega^2+\G^2\right)^2}\right)\right]\right\},~~\Omega = \hbar\omega.\label{zero}
\end{align}
In the limit $\Gamma \rightarrow \infty$  the zero-temperature PSD thus scales like $\tilde{S}^{\rm h}(\omega) \propto \omega^3$. This is in full agreement with the results obtained in  Refs. \cite{averin10prl, sergi11prb}, where
this asymptotic behavior has been found as being uniform throughout the whole frequency region. However, this is no longer the case when $\Gamma$
is finite: the second term in the rhs of Eq.~\eqref{zero} introduces a linear cutoff in the limit $\omega \rightarrow 0$, $\tilde{S}^{\rm h}(\omega) \propto \omega$.  In distinct contrast, in the high-frequency region, the first term in the rhs of Eq.~\eqref{zero} is dominating. As a result  the PSD~\eqref{zero} approaches a square-law asymptotic dependence,  $\tilde{S}^{\rm h}(\omega) \propto \omega^2$,  in the high-frequrncy limit, see Fig.~\ref{fig:ze}.

\section{Conclusions}
By using the Green function formalism we have investigated electronic heat  transport with our focus being the heat flow fluctuations for a setup composed of a single orbital molecular wire. For the noninteracting case we succeeded in deriving a closed form for the frequency dependence of heat current noise, i.e., the heat noise PSD, see Eq.~\eqref{longeqn},  both in nonequilibrium $T_{\rm L} \neq T_{\rm R}$ and  in thermal equilibrium $T_{\rm L}=T_{\rm R}$. The dependence of the heat noise on the orbital energy $\varepsilon_0$ is qualitatively  different from that for the accompanying electric current noise. 

In the zero-temperature limit, the PSD of the heat noise obeys two distinctive asymptotic behaviors, being different in the low-frequency and the high-frequency regions. It is evident that the particular square-law shape of the PSD in the high-frequency region is due to the Lorentzian shape of the transmission coefficient, $\ma{T}(E)$. Yet the general effect would remain  for any choice of the coefficient in the form of a localized, bell-shaped function: the noise spectrum will deviate from a cubic power-law asymptotic behavior upon entering the high-frequency region.

There is  an intriguing perspective to apply an  external periodic perturbation  with the goal to control the spectral properties of the heat noise, similar to electronic shot-noise control in ac-driven nanoscale conductors \cite{camalet04prb}. This idea can be realized, for example, by subjecting the molecular wire to strong laser radiation \cite{kohler07naturenano} or by using direct modulations of the gate voltage. We conjecture that the role of laser radiation may give rise to novel phenomena to be explored further  by combining a   Floquet theory for the driven system with the nonequilibrium Green function formalism \cite{camalet04prb,kohler05,rey07prb}.

As emphasized in our introduction, with this work only the electron subsystem has been considered. Realistic heat transport  in  real molecular junctions  would involve the complexity of interacting electrons and electron-phonon interactions \cite{dubi11rmp}. As reasoned in the introduction, the electronic heat transport may dominate in certain situations so that the measured heat noise can be attributed approximately to the electronic component only.  The unified approach, which would include both  the electron and the phonon subsystems, as well as the effects of their interactions, presents a future challenge, although  several contributions in this direction for the heat current (but not the heat PSD)  have already been undertaken before  \cite{dubi11rmp,galperin07prb,galperin07jpcm}).

\begin{acknowledgments}
The work has been supported by the German Excellence
Initiative via the ``Nanosystems Initiative Munich'' (NIM) (P.H.) and the DFG priority
program DFG-1243 ``Quantum transport at the molecular scale'' (F.Z.,
P.H.).
\end{acknowledgments}

\end{document}